\documentclass[journal]{IEEEtran}

\usepackage{cite}
\usepackage{graphicx}
\usepackage{balance}
\usepackage{color}
\usepackage{url}
\usepackage{graphicx, pstool}
\usepackage{float}
\usepackage{array}
\usepackage[english]{babel} 
\usepackage{amsmath,amssymb,amsfonts}
\usepackage{textcomp}
\usepackage{algorithmic}
\usepackage{caption}
\usepackage{subcaption}
\usepackage{breqn}
\usepackage{blindtext}
\usepackage{dblfloatfix} 
\usepackage[ruled,vlined,linesnumbered]{algorithm2e}
\usepackage{multirow}
\makeatletter
\renewcommand{\fnum@figure}{Fig. \thefigure}
\makeatother

\newcolumntype{P}[1]{>{\centering\arraybackslash}p{#1}}
\newtheorem{theorem}{Theorem}

\newtheorem{lemma}{Lemma}

\newtheorem{corollary}{Corollary}

\newtheorem{proposition}{Proposition}

\newtheorem{conjecture}{Conjecture}

\newtheorem{sketch}{Sketch of Proof}

\IEEEaftertitletext{\vspace{-1\baselineskip}}

\begin{document}
	
	\title{OTFS based Joint Radar and Communication: Signal Analysis using the Ambiguity Function}
	\author{
		Shalanika Dayarathna,~\IEEEmembership{Member,~IEEE,} Peter Smith,~\IEEEmembership{Fellow,~IEEE,} \\ Rajitha Senanayake,~\IEEEmembership{Member,~IEEE,} and Jamie Evans~\IEEEmembership{Senior Member,~IEEE}
	}	
	\maketitle
	
	\begin{abstract}
		Orthogonal time frequency space (OTFS) modulation has recently been identified as a suitable waveform for joint radar and  communication systems. Focusing on the effect of data modulation on the radar sensing performance, we derive the ambiguity function (AF) of the OTFS waveform and characterize the radar global accuracy. We evaluate the behavior of the AF with respect to the distribution of the modulated data and derive an accurate approximation for the mean and variance of the AF, thus, approximating its distribution by a Rice distribution. Finally, we evaluate the global radar performance of the OTFS waveform with the OFDM waveform. 
	\end{abstract}
	
	
	\section{Introduction}\label{Sec-Intro}
	The orthogonal time frequency space (OTFS) waveform has recently been identified as a promising candidate for emerging joint radar and communication (JRC) systems \cite{2998583}. Compared to orthogonal frequency division multiplexing (OFDM), OTFS modulation is less susceptible to extreme Doppler channels, and thus achieves significantly better performance  in high mobility networks such as high speed trains, unmanned aerial vehicles (UAVs) and vehicular ad hoc networks \cite{3129975,2000408}.
	
	The feasibility of using the OTFS waveform for radar systems is analyzed in \cite{8835764}, where it is shown that OTFS achieves longer range, faster tracking rates and detection of higher Doppler frequencies compared to OFDM based radar. Taking a step further, a JRC system is considered in \cite{2998583, 9266546}, where the Cramér-Rao Lower Bounds (CRLBs) for range and velocity estimation using the maximum likelihood (ML) detection are derived under the OTFS waveform. Furthermore, focusing on the mean squared error (MSE) of range and velocity estimation, the radar performance of OTFS is compared with that of OFDM and the frequency modulated continuous wave radar waveform. In \cite{9473534}, a generalized likelihood ratio test based estimator that utilizes inter-carrier interference and inter-symbol-interference is proposed to surpass the maximum unambiguous detection limits in range and velocity. In \cite{9838274}, a discrete Fourier transform spread OTFS scheme is proposed to reduce the peak-to-average power ratio (PAPR) while achieving super resolution sensing accuracy.
	
	While much research has investigated the performance of radar under OTFS modulation, the focus of the analytical work has been limited to local accuracy measures such as CRLB and MSE or power efficiency measures such as PAPR and out-of-band (OoB) power radiation. However, the radar global accuracy, which is characterised by the behavior of sidelobes in the ambiguity function (AF) \cite{2785598,01611-4,7507195} has gained much less research attention. Some limited work has focused on the AF of OTFS waveforms but the results are limited to simulations \cite{9414107,9473601,9904977,9904976}. The behavior of the AF is analyzed in \cite{9259515} for an OFDM based JRC system. To the best of our knowledge, such analysis has not been extended to OTFS based JRC systems. 
	
	In this letter, we address this gap by providing a thorough analysis of the AF for modulated OTFS used in JRC systems. Such an analysis is extremely useful in characterizing the general radar performance and for the process of designing radar friendly coding schemes for OTFS signals in JRC systems. By considering two radar performance metrics, namely the peak-sidelobe ratio (PSLR) and integrated sidelobe ratio (ISLR), we show that the global radar performance of OTFS is highly variant according to the modulated data. Next, we evaluate the behavior of the AF with respect to the distribution of the transmitted communication symbols and derive an accurate approximation for the mean and variance of the AF, thus, enabling the global evaluation of the general radar performance in OTFS averaged over all possible data sequences. 
	
	\section{Structure of OTFS Signal}\label{Sec-Model}
	In OTFS, the communication data is modulated in the delay-Doppler (DD) domain in contrast to the traditional time-frequency (TF) domain modulation that is used, for example, in OFDM. Similar to OFDM, with the new OTFS modulation, for a given symbol duration $T$ and frequency separation $\Delta f=1/T$, $NM$ communication symbols can be modulated over a bandwidth of $M\Delta f$ and a time interval, $NT$ where $N$ and $M$ are positive integers. However, in contrast to OFDM, the orthogonality between communication symbols is achieved in the DD-domain while symbols overlap in the TF-domain. This is achieved by modulating data symbols on to two-dimensional sinc functions separated by roughly $T/M$ along the delay domain and $\Delta f/N$ along the Doppler domain. The resulting OTFS signal can be expressed as \cite{3069913},
	\begin{align}\label{eq_1}
		s(t) = \dfrac{1}{\sqrt{NM}}\sum_{n=0}^{N-1}\sum_{m=0}^{M-1} X[n,m] g(t-nT) \textrm{e}^{j2\pi m\Delta f(t-nT)},
	\end{align}
	where $g(t)$ is the rectangular pulse with amplitude $1/\sqrt{T}$ in $0 \le t < T$ and otherwise zero, and $X[n,m]$ denotes the $(nm)$-th element of the TF-domain equivalent symbol matrix. When $x[k,l]$ denotes the $(kl)$-th element in the modulated symbol matrix in the DD-domain, which is a complex value drawn from the symbol set of $M^{'}$-QAM, $X[n,m]$ is expressed as, 	
	\begin{align}
		X[n,m] &= \sum_{k=0}^{N-1}\sum_{l=0}^{M-1} x[k,l]\textrm{e}^{j2\pi \bigg(\dfrac{nk}{N}-\dfrac{ml}{M}\bigg)}. \label{eq_2} 
	\end{align}
		
	\begin{figure*}[b]
		\begin{subfigure}{.5\textwidth}
			\centering
			\includegraphics[width=\textwidth]{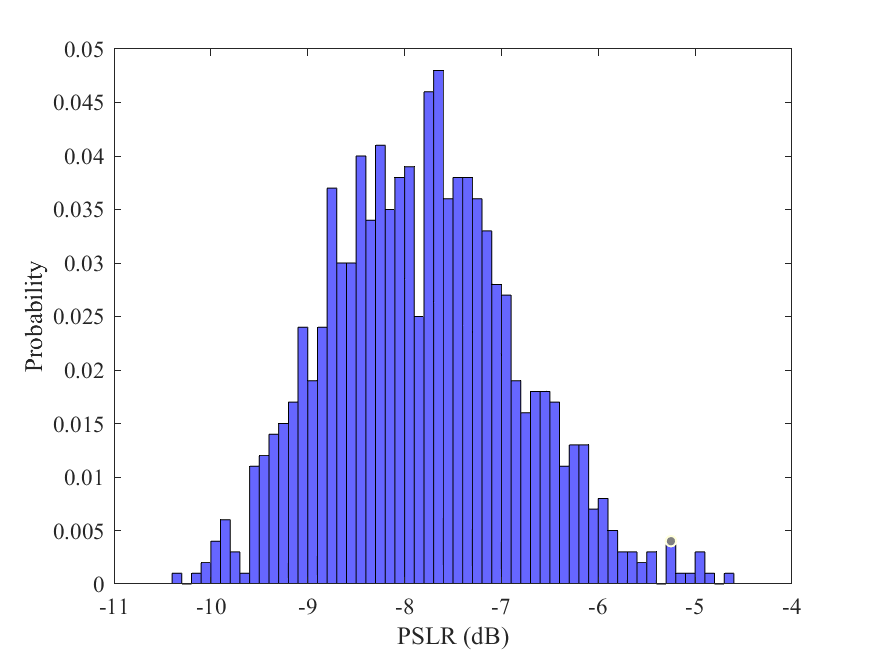}
			\captionsetup{justification=centering}
			\caption{The PSLR distribution}
		\end{subfigure}
		\begin{subfigure}{.5\textwidth}
			\centering
			\includegraphics[width=\textwidth]{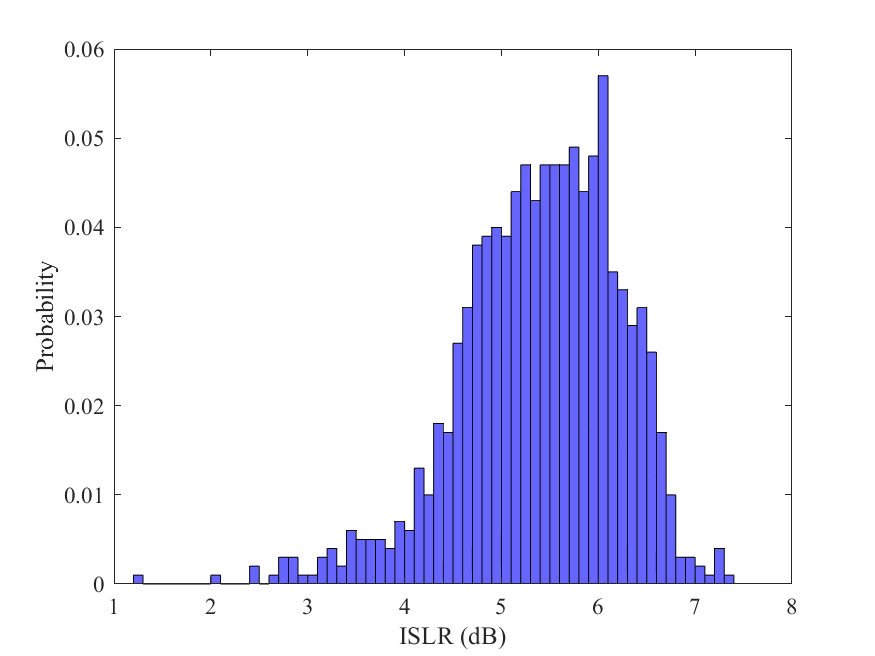}
			\captionsetup{justification=centering}
			\caption{The ISLR distribution}
		\end{subfigure}
		\caption{The sidelobe behavior of 4-QAM modulated OTFS with $N=4, M=4$.}
		\label{figure1}
	\end{figure*}
	\subsection{Communication Performance of OTFS Waveform}
	The communication performance of the OTFS waveform has been thoroughly discussed in the literature. Thus, in this work, we have limited our scope to analyze the the radar performance. The communication performance of the OTFS modulation in high mobility applications is discussed in detail in \cite{Monk2016OTFSORTHOGONALTF}. In the context of JRC, the communication performance of OTFS in terms of pragmatic capacity \cite{2998583} and bit error rate (BER) \cite{3117404} is shown to be higher than that of OFDM.
	
	\section{Radar Performance of OTFS Waveform}\label{Sec-Radar}
	A received radar signal undergoes a delay and Doppler shift associated with the range and the velocity of the target. The AF represents the time response of a matched filter when the transmitted signal is received with a certain time delay, $\tau$, and a Doppler shift, $f_d$. Therefore, the AF is an important tool to analyze target estimates and design radar waveforms \cite{0471663085}. 
	\setcounter{equation}{2}
	\begin{figure*}	
		\begin{align}
			\hat{A}(\tau,f_d) \!=\! &\dfrac{1}{NM}\sum_{n_1=0}^{N-1}\sum_{n_2=0}^{N-1}\sum_{m_1=0}^{M-1}\sum_{m_2=0}^{M-1}\, \biggl[ X[n_1,m_1]X^{*}[n_2,m_2]\textrm{e}^{j\pi (m_1+m_2)\Delta f\tau}\textrm{e}^{j\pi [(n_1+n_2+1)T+\tau]f_d} \nonumber \\ & \hspace{250pt} \times A_g\bigg([n_1-n_2]T-\tau,[m_1-m_2]\Delta f+f_d\bigg)\biggr]. \label{eq_8} 
		\end{align}	
		\hrule
	\end{figure*}
	The complex AF of modulated OTFS can be expressed as \eqref{eq_8}, given at the top of the next page, where for notational convenience $A_g(t^{'},\alpha)$ is defined as 
	\setcounter{equation}{3}
	\begin{align}\label{eq_9}
		A_g(t^{'},\alpha) \!=\! \left\{
		\begin{array}{ll}
			\!\!\!\! \dfrac{(T\!-\!|t^{'}|)}{T}\, \textrm{sinc}(\pi \alpha(T\!-\!|t^{'}|)) & \quad |t^{'}|<T \\
			0 & \textrm{otherwise.}
		\end{array}
		\right. 
	\end{align}	
	Then, the AF, $A(\tau,f_d)$, can be computed as $|\hat{A}(\tau,f_d)|$. Noting that the local radar performance of OTFS is well-studied, we focus on two global radar performance metrics.
	
	\subsection{Peak-to-Sidelobe Ratio (PSLR)}	
	The PSLR is a key performance metric used to measure sidelobe behaviour of the AF. It represents the largest sidelobe to the mainlobe peak as given below \cite{7507195},
	\begin{align}\label{eq_3.1}
		\textrm{PSLR} = 20\log\bigg(\dfrac{|\hat{A}(\bar{\tau},\bar{f_d})|}{|\hat{A}(0,0)|}\bigg),
	\end{align}
	where $(\bar{\tau},\bar{f_d})$ is the location of the largest sidelobe outside the mainlobe region. The PSLR is used to identify the capability of detecting weak targets in the presence of nearby interfering targets. Smaller PSLR indicates a smaller probability of false alarm and is a desired property of radar systems. 
	
	\subsection{Integrated Sidelobe Ratio (ISLR)}
	The ISLR evaluates the ratio of the total sidelobe energy to the energy of the mainlobe in the AF and is given by \cite{7507195},		
	\begin{align}\label{eq_3.2}
		\textrm{ISLR} \!=\! 10\log\!\bigg(\dfrac{\int_{\tau=-NT}^{NT}\int_{f_d=-\infty}^{\infty}|\hat{A}(\tau,f_d)|^2d\tau \,df_d}{\int_{\tau=-T}^{T} \int_{f_d=-\Delta f}^{\Delta f}|\hat{A}(\tau,f_d)|^2 d\tau d\,f_d}\!-\!1\!\bigg).
	\end{align}
	The ISLR measures how much energy is leaking from the mainlobe. As such, it is very important in dense target sensing where distributed clutter is present.
		
	Fig.~\ref{figure1} plots the distribution of the PSLR and ISLR of OTFS over $1000$ realizations with $N=4, M=4$ and random communication data. From the plot, we observe that the PSLR varies from $-10.3$ to $-4.6$ while the ISLR varies from $1.2$ to $7.4$ based on the modulated communication data. Therefore, the sidelobe behavior depends on the communication data modulated onto the OTFS signal. As such, the AF obtained for the OTFS signal modulated with one random modulated data sequence cannot be used to evaluate or to compare the radar performance of the modulated OTFS waveform with other traditional radar waveforms. Therefore, in the following we evaluate the behavior of the AF for modulated OTFS signals based on statistical properties.
	
	\setcounter{equation}{6}
	\begin{figure*}
		\begin{align}\label{eq_16} 
			E\{A^2(\tau,f_d)\} &\!=\! N^2M^2|A_g(-\tau,f_d)|^2C_1(Tf_d,N)C_1(\Delta f\tau,M) \!+\! \sum_{n_1=0}^{N-1}\sum_{n_2=0}^{N-1} \sum_{m_1=0}^{M-1}\sum_{m_2=0}^{M-1} \bigg|A_g\bigg([n_1\!-\!n_2]T\!-\!\tau,[m_1\!-\!m_2]\Delta f\!+\!f_d\bigg)\!\bigg|^2 \nonumber \\ 
			&\qquad + \sum_{n_1=0}^{N-1}\sum_{n_2=0}^{N-1}\sum_{\bar{n}_1=0}^{N-1}\sum_{\bar{n}_2=0}^{N-1}\sum_{m_1=0}^{M-1}\sum_{m_2=0}^{M-1}\sum_{\bar{m}_1=0}^{M-1}\sum_{\bar{m}_2=0}^{M-1} \biggl\{ \textrm{e}^{j\pi (m_1+m_2-\bar{m}_1-\bar{m}_2)\Delta f\tau}\textrm{e}^{j\pi (n_1+n_2-\bar{n}_1-\bar{n}_2)Tf_d} \nonumber \\ & \qquad A_g\bigg([n_1-n_2]T-\tau,[m_1-m_2]\Delta f+f_d\bigg) A_g^{*}\bigg([\bar{n}_1-\bar{n}_2]T-\tau,[\bar{m}_1-\bar{m}_2]\Delta f+f_d\bigg) \biggl[C_2\bigg(\dfrac{n_1+\bar{n}_2}{N}\bigg)\nonumber \\ & \qquad \quad C_2\bigg(\dfrac{n_2+\bar{n}_1}{N}\bigg) C_2\bigg(\dfrac{m_1+\bar{m}_2}{M}\bigg)C_2\bigg(\dfrac{m_2+\bar{m}_1}{M}\bigg)|E\{x^2[k,l]\}|^2 + \dfrac{1}{NM}C_2\bigg(\dfrac{n_1-n_2-\bar{n}_1+\bar{n}_2}{N}\bigg)\nonumber \\ & \qquad \qquad C_2\bigg(\dfrac{m_1-m_2-\bar{m}_1+\bar{m}_2}{M}\bigg) \bigg(E\{|x[k,l]|^4\} - |E\{x^2[k,l]\}|^2 - 2\bigg) \biggr] \biggr\}.
		\end{align}
		\hrule
	\end{figure*}
	\section{Ambiguity Function Analysis}\label{Sec-AF}
	In this section, we model the behavior of the AF in \eqref{eq_8} by computing the mean, $E\{A(\tau,f_d)\}$, and the variance, $\textrm{Var}\{A(\tau,f_d)\}$, over the modulated data. As the computation of variance requires both first and second moments of $A(\tau,f_d)$, we focus on deriving $E\{A^2(\tau,f_d)\}$ and $E\{A(\tau,f_d)\}$. 
	
	Motivated by the randomness associated with the communication data, we assume independent and identically distributed (i.i.d) modulated symbols \cite{2927326}. Thus, with normalized symbol energy $E\{x[k_1,l_1]x^{*}[k_2,l_2]\}$ is one when $k_1=k_2$, $l_1=l_2$ and zero otherwise. Writing $\textrm{e}^{j2\pi n}=1, \forall n\in \mathbb{Z}$, $E\{A^2(\tau,f_d)\} = E\{\hat{A}(\tau,f_d)\hat{A}^{*}(\tau,f_d)\}$, and the linearity of the expectation, $E\{A^2(\tau,f_d)\}$ is computed as \eqref{eq_16}, given at the top of the page, where	
	\setcounter{equation}{7}
	\begin{align}	
		C_1(x,N) &= \left\{
		\begin{array}{ll}
			1 & x\in \mathbb{Z} \\
			\textrm{sinc}^2(N\pi x)/\textrm{sinc}^2(\pi x)\qquad & \textrm{otherwise},
		\end{array}
		\right. \label{eq_C1}\\
	\end{align}
	\begin{align}
		C_2(x) &= \left\{
		\begin{array}{ll}
			1 \qquad\qquad\qquad\qquad\quad\; & x\in \mathbb{Z} \\
			0 & \textrm{otherwise}.
		\end{array}
		\right. \label{eq_C2}
	\end{align}
	Please refer to Appendix \ref{appendixA} for the full derivation of \eqref{eq_16}. We also note that in \eqref{eq_16}, only $E\{|x[k,l]|^4\}$ and $|E\{x^2[k,l]\}|^2$ depend on $M^{'}$. 
	
	Computation of $E\{A(\tau,f_d)\}$ requires the expectation of the absolute value of a summation which is mathematically  intractable. Thus, we  use Jensen's inequality for the concave square root function and for the convex absolute value function and derive bounds for $E\{A(\tau,f_d)\}$ and $\textrm{Var}\{A(\tau,f_d)\}$ as,
	\begin{align}
		|E\{\hat{A}(\tau,f_d)\}| &\le E\{A(\tau,f_d)\} \le \sqrt{E\{A^2(\tau,f_d)\}}, \label{eq_11} \\
		0 &\le \textrm{Var}\{A(\tau,f_d)\} \le 2\sigma^2, \label{eq_12}
	\end{align}
	where $\sigma^2 =  \big(E\{A^2(\tau,f_d)\}-|E\{\hat{A}(\tau,f_d)\}|^2\big)/2$ with $|E\{\hat{A}(\tau,f_d)\}|$ given in \eqref{eq_15} at the top of the next page. Please refer to Appendix \ref{appendixB} for the full derivation of \eqref{eq_15}.
	\setcounter{equation}{11}
	\begin{figure*}
		\begin{align}\label{eq_15}
			|E\{\hat{A}(\tau,f_d)\}| &= \left\{
			\begin{array}{ll}
				NM|\textrm{sinc}(\pi NTf_d)| & \tau=0 \\
				\dfrac{NM}{T}(T-|\tau|)\bigg|\dfrac{\textrm{sinc}(\pi M\Delta f\tau)\,\textrm{sinc}(\pi [T-|\tau|]f_d)}{\textrm{sinc}(\pi\Delta f\tau)}\bigg| & 0<|\tau|<T, Tf_d\in \mathbb{Z} \\
				\dfrac{NM}{T}(T-|\tau|)\bigg|\dfrac{\textrm{sinc}(\pi M\Delta f\tau)\,\textrm{sinc}(\pi NTf_d)\,\textrm{sinc}(\pi [T-|\tau|]f_d)}{\textrm{sinc}(\pi\Delta f\tau)\,\textrm{sinc}(\pi Tf_d)}\bigg| \qquad \qquad  & 0<|\tau|<T, Tf_d\not\in \mathbb{Z}\\
				0 & |\tau| \ge T
			\end{array}
			\right.
		\end{align}
		\hrule
	\end{figure*}	
	
	The complex AF in \eqref{eq_8} is a summation of a large number of random variables driven by i.i.d data symbols. Whilst not shown here due to page limitations, motivated by the central limit theorem, for large $N$ and $M$ values, the distribution of $\hat{A}(\tau,f_d)$ can be approximated using a complex Gaussian distribution. Thus, $A(\tau,f_d)$ can be approximated by a Rice distribution where the mean and variance only depend on $E\{A^2(\tau,f_d)\}$ and $E\{\hat{A}(\tau,f_d)\}$. Next, using known results for the Rice distribution, a reasonable approximation for $E\{A(\tau,f_d)\}$ and $\textrm{Var}\{A(\tau,f_d)\}$ can be derived as,
	\setcounter{equation}{12}
	\begin{align}
		E\{A(\tau,f_d)\} &\!\approx\! \sigma\sqrt{\pi/2}\textrm{L}_{1/2} \bigg(\dfrac{-|E\{\hat{A}(\tau,f_d)\}|^2}{2\sigma^2}\bigg), \label{eq_13}
	\end{align}
	\begin{align}
		\textrm{Var}\{\!A(\tau,f_d)\} &\!\approx\! E\{A^2(\tau,f_d)\} \!-\! \dfrac{\pi \sigma^2}{2} \textrm{L}_{1/2}^2 \bigg(\!\!\dfrac{-|E\{\!\hat{A}(\tau,f_d)\!\}|^2}{2\sigma^2}\!\bigg)\!, \label{eq_14}
	\end{align}
	where $L_{1/2}(.)$ is the Laguerre function \cite{0524047}. 	
	
	\section{Numerical Results}
	In this section, we evaluate the accuracy of the approximations derived in section \ref{Sec-AF} and use the approximated AF to evaluate the sidelobe behavior of the OTFS waveform with respect to existing JRC waveforms such as OFDM.
	
	Table \ref{table1} presents the relative error of the upper bounds in \eqref{eq_11} and \eqref{eq_12} and the approximations in \eqref{eq_13} and \eqref{eq_14} with respect to the simulated values of $E\{A(\tau,f_d)\}$ and $\textrm{Var}\{A(\tau,f_d)\}$ averaged over the $(\tau,f_d)$ coordinates. The simulated $E\{A(\tau,f_d)\}$ and $\textrm{Var}\{A(\tau,f_d)\}$ values are computed for $N=4,M=8, |\tau|\le NT$ and $|f_d|\le10$ by taking the mean and variance of $A(\tau,f_d)$ over $1000$ realizations of $M^{'}$-QAM where $M^{'}=4,8,16,32$ and $64$. 
	From the table, we observe that for $E\{A(\tau,f_d)\}$ and $\textrm{Var}\{A(\tau,f_d)\}$, the average relative error of the upper bound is over $4.5$ and $20$ times higher than the approximation, respectively. We can also observe that the accuracy of the approximation is independent of $M^{'}$. Note that obtaining the exact value of $\textrm{Var}\{A(\tau,f_d)\}$ requires the computation of the AF for all possible modulated data sequences. The complexity of this computation increases exponentially with $M$ and $N$. Due to this and the above 300\% relative error in \eqref{eq_12}, the approximation error of 13\% in \eqref{eq_14} can be considered acceptable. As such, the approximations in \eqref{eq_13} and \eqref{eq_14} can be used to evaluate the behavior of the AF and to compare the global accuracy of radar sensing of the modulated OTFS signals with other waveforms used in radar sensing and JRC systems. 
	\begin{table}
		\centering
		\caption{Average relative error in computing $E\{A(\tau,f_d)\}$ and $\textrm{Var}\{A(\tau,f_d)\}$ over $(\tau,f_d)$ coordinates}		
		\renewcommand{\arraystretch}{1.2}
		\begin{tabular}{c|c|c|c|c|}
			\cline{2-5}
			\rule{0pt}{10pt}
			& \multicolumn{2}{c|}{$E\{A(\tau,f_d)\}$} & \multicolumn{2}{c|}{$\textrm{Var}\{A(\tau,f_d)\}$} \\
			\cline{2-5}
			& Approx. & Upper bound & Approx. & Upper bound \\
			\hline	 				
			\multicolumn{1}{|c|}{4-QAM} & 0.023 & 0.145 & 0.106 & 3.3273 \\ \hline
			\multicolumn{1}{|c|}{8-QAM} & 0.029 & 0.154 & 0.133 & 3.040 \\ \hline 
			\multicolumn{1}{|c|}{16-QAM} & 0.021 & 0.140 & 0.128 & 3.070 \\ \hline 
			\multicolumn{1}{|c|}{32-QAM} & 0.023 & 0.145 & 0.129 & 3.087 \\ \hline 
			\multicolumn{1}{|c|}{64-QAM} & 0.025 & 0.152 & 0.132 & 3.082 \\ \hline 
		\end{tabular}
		\label{table1}
	\end{table}
	
\begin{figure}
		\centering\includegraphics[width=0.5\textwidth]{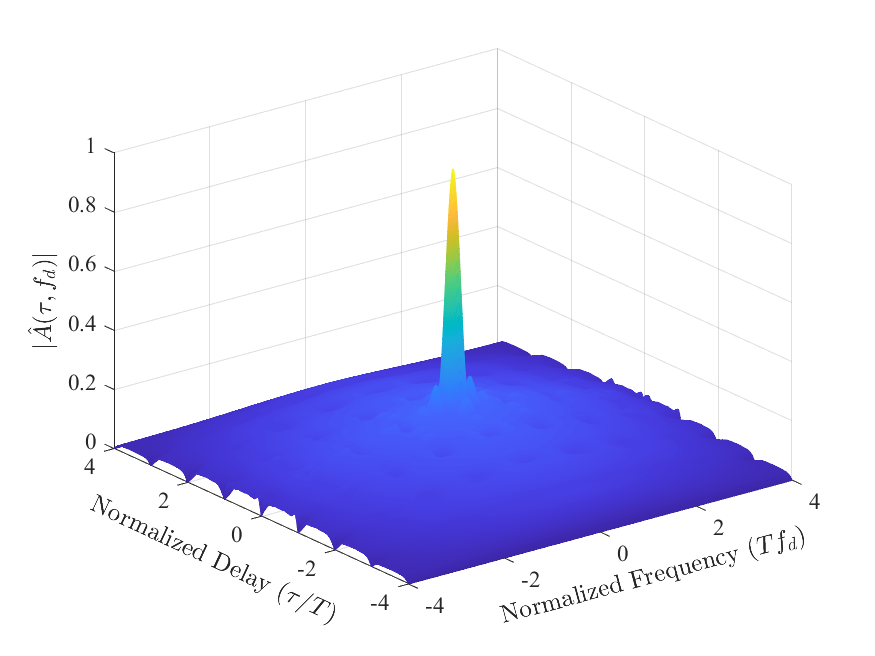}
		\captionsetup{justification=centering}
		\caption{Approximated OTFS AF using $E\{A(\tau,f_d)\}$}
		\label{figure2}
	\end{figure}

	Fig.~\ref{figure2} plots the average AF for an i.i.d symbol distribution when $M=N=4$ with $4$-QAM using the approximation in \eqref{eq_13}. We observe that on average the sidelobes of the OTFS waveform are negligible, suggesting that the global accuracy of radar sensing using the OTFS waveform with an i.i.d symbol distribution is acceptable in JRC systems.
	
\begin{figure}
		\centering\includegraphics[width=0.5\textwidth]{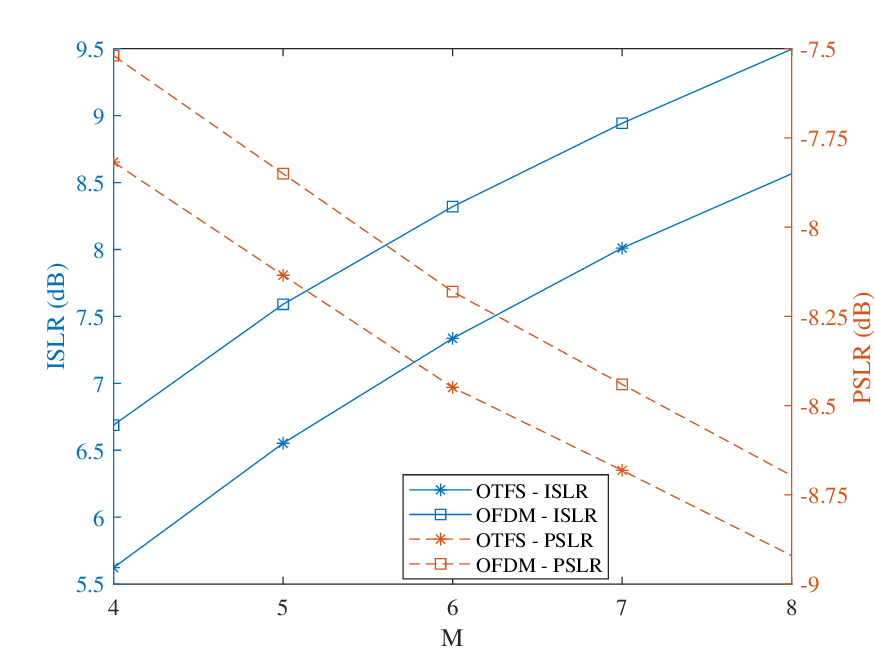}
		\captionsetup{justification=centering}
		\caption{ISLR and PSLR of OTFS vs OFDM with $M$}
		\label{figure3}
	\end{figure}

	Next, we compare the effectiveness of the modulated OTFS and OFDM for radar sensing using ISLR and PSLR computed from the magnitude of the complex AF in \eqref{eq_8}. We note that the complex AF of OFDM can be obtained using \eqref{eq_8} with $X[n,m]=x[n,m], \forall n,m$. Fig.~\ref{figure3} plots the average ISLR and PSLR values for the  modulated OTFS and OFDM versus $M$ over $1000$ realizations with $N=4$ when the mainlobe region is defined as $|\tau|<T$ and $|f_d|<\Delta f$. As the energy of the waveforms are kept constant, the height of sidelobe peaks decreases with $M$ resulting in a reduction of PSLR. On the other hand, with increasing $M$, the delay resolution increases, thus, decreasing the mainlobe width when $f_d=0$. As such, more energy leaks into the sidelobe region, thus increasing ISLR with $M$. We note that large $M$ and $N$ values are required to achieve a higher communication data rate by increasing the number of data symbols that can be modulated into the OTFS waveform. As such, it is important to select $M$ based on the importance of the ISLR and the delay resolution for the considered radar application as well as considering the trade-off between the ISLR and the communication data rate. From the plot, we can observe that on average modulated OTFS has better radar performance in terms of both ISLR and PSLR compared to modulated OFDM signals.
	
	\section{Conclusion}\label{Sec-Con}
	We focused on the use of the new OTFS waveform for JRC and analyzed its AF to characterize the global accuracy of radar sensing when the signal is modulated with communication data. Considering the QAM-modulated OTFS signal, we showed that the sidelobe structure depends on the distribution of the communication symbols. We also derived an accurate approximation on the mean and variance of the AF under the assumption of an i.i.d symbol distribution and showed that the distribution of the AF can be approximated by a Rice distribution at each DD coordinate. 
	
	\begin{figure*}
		\begin{align}
			E\{A^2(\tau,f_d)\} &\!=\! \dfrac{1}{N^2M^2}\!\!\sum_{n_1=0}^{N-1}\sum_{n_2=0}^{N-1}\sum_{m_1=0}^{M-1}\sum_{m_2=0}^{M-1}	\sum_{\bar{n_1}=0}^{N-1}\sum_{\bar{n_2}=0}^{N-1} \sum_{\bar{m_1}=0}^{M-1}\sum_{\bar{m_2}=0}^{M-1}\biggl[\textrm{e}^{j\pi (m_1+m_2-\bar{m_1}-\bar{m_2})\Delta f\tau}\textrm{e}^{j\pi [n_1+n_2-\bar{n_1}-\bar{n_2}]Tf_d} \nonumber \\ & \, 
			E\{X[n_1,m_1]X^{*}[n_2,m_2]X^{*}[\bar{n_1},\bar{m_1}]X[\bar{n_2},\bar{m_2}]\} \times A_g\bigg([n_1-n_2]T-\tau,[m_1-m_2]\Delta f+f_d\bigg) \nonumber \\ &
			\hspace{190pt} \times A_g^{*}\bigg([\bar{n_1}-\bar{n_2}]T-\tau,[\bar{m_1}-\bar{m_2}]\Delta f+f_d\bigg)\biggr]. \label{eq_app7} 
		\end{align}
		\hrule
	\end{figure*}
	\setcounter{equation}{16}
	\begin{figure*}
		\begin{align}\label{eq_app10} 
			&E\{X[n_1,m_1]X^{*}[n_2,m_2]X^{*}[\bar{n_1},\bar{m_1}]X[\bar{n_2},\bar{m_2}]\} \nonumber \\ &\hspace{80pt} = E\{|x[k,l]|^4\} \biggl( \sum_{k_1=0}^{N-1} \textrm{e}^{\frac{j2\pi k_1}{N} \big(n_1-n_2-\bar{n_1}+\bar{n_2}\big)} \sum_{l_1=0}^{M-1} \textrm{e}^{\frac{-j2\pi l_1}{M} \big(m_1-m_2-\bar{m_1}+\bar{m_2}\big)}\biggr) + \nonumber \\ &\hspace{100pt} |E\{|x[k,l]|^2\}|^2 \biggl( - 2 \sum_{k_1=0}^{N-1} \textrm{e}^{\frac{j2\pi k_1}{N} \big(n_1-n_2-\bar{n_1}+\bar{n_2}\big)} \sum_{l_1=0}^{M-1}\textrm{e}^{\frac{-j2\pi l_1}{M} \big(m_1-m_2-\bar{m_1}+\bar{m_2}\big)} \nonumber \\ &\hspace{120pt} + \sum_{k_1=0}^{N-1} \textrm{e}^{\frac{j2\pi k_1}{N} \big(n_1-n_2\big)} \sum_{\bar{k_1}=0}^{N-1} \textrm{e}^{\frac{-j2\pi \bar{k_1}}{N} \big(\bar{n_1}-\bar{n_2}\big)} \sum_{l_1=0}^{M-1} \textrm{e}^{\frac{-j2\pi l_1}{M} \big(m_1-m_2\big)} \sum_{\bar{l_1}=0}^{M-1} \textrm{e}^{\frac{j2\pi \bar{l_1}}{M} \big(\bar{m_1}-\bar{m_2}\big)} \nonumber \\ &\hspace{120pt} + \sum_{k_1=0}^{N-1} \textrm{e}^{\frac{j2\pi k_1}{N} \big(n_1-\bar{n_1}\big)} \sum_{k_2=0}^{N-1} \textrm{e}^{\frac{-j2\pi k_2}{N} \big(n_2-\bar{n_2}\big)} \sum_{l_1=0}^{M-1} \textrm{e}^{\frac{-j2\pi l_1}{M} \big(m_1-\bar{m_1}\big)} \sum_{l_2=0}^{M-1} \textrm{e}^{\frac{j2\pi l_2}{M} \big(m_2-\bar{m_2}\big)} \biggr) + \nonumber \\ &\hspace{100pt} |E\{x^2[k,l]\}|^2 \biggl(- \sum_{k_1=0}^{N-1} \textrm{e}^{\frac{j2\pi k_1}{N} \big(n_1-n_2-\bar{n_1}+\bar{n_2}\big)} \sum_{l_1=0}^{M-1} \textrm{e}^{\frac{-j2\pi l_1}{M} \big(m_1-m_2-\bar{m_1}+\bar{m_2}\big)} \nonumber \\ &\hspace{120pt} + \sum_{k_1=0}^{N-1} \textrm{e}^{\frac{j2\pi k_1}{N} \big(n_1+\bar{n_2}\big)} \sum_{k_2=0}^{N-1} \textrm{e}^{\frac{-j2\pi k_2}{N} \big(n_2+\bar{n_1}\big)} \sum_{l_1=0}^{M-1} \textrm{e}^{\frac{-j2\pi l_1}{M} \big(m_1+\bar{m_2}\big)} \sum_{l_2=0}^{M-1} \textrm{e}^{\frac{j2\pi l_2}{M} \big(m_2+\bar{m_1}\big)} \biggr).
		\end{align}
		\hrule
	\end{figure*}
	\begin{appendices}		
		\section{}\label{appendixA}	
		Let us start by writing $A^2(\tau,f_d) = |\hat{A}(\tau,f_d)|^2 = \hat{A}(\tau,f_d)\hat{A}^{*}(\tau,f_d)$. Therefore, $E\{A^2(\tau,f_d)\}$ can be written as \eqref{eq_app7}, given at the top of the next page. In order to compute $E\{A^2(\tau,f_d)\}$, we first consider the i.i.d structure of $x[k,l]$ and write 
		\setcounter{equation}{15}
		\begin{align}\label{eq_app9} 
			&E\{x[k_1,l_1]x^{*}[k_2,l_2]x^{*}[\bar{k_1},\bar{l_1}]x[\bar{k_2},\bar{l_2}]\} \nonumber \\ &=\! \left\{ \begin{array}{ll}
				\hspace{-8pt}E\{|x[k_1,l_1]|^4\} & k_1\!=\! k_2\!=\!\bar{k_1}\!=\!\bar{k_2}, l_1\!=\! l_2\!=\!\bar{l_1}\!=\!\bar{l_2} \\
				\hspace{-8pt}|E\{|x[k_1,l_1]|^2\}|^2 & k_1\!=\! k_2,\bar{k_1}\!=\!\bar{k_2}, l_1\!=\! l_2,\bar{l_1}\!=\!\bar{l_2} \textrm{ except } \\ &  k_1\!=\! k_2\!=\!\bar{k_1}\!=\!\bar{k_2}, l_1\!=\! l_2\!=\!\bar{l_1}\!=\!\bar{l_2}\\
				\hspace{-8pt}|E\{|x[k_1,l_1]|^2\}|^2 & k_1\!=\!\bar{k_1}, k_2\!=\!\bar{k_2}, l_1\!=\!\bar{l_1}, l_2\!=\!\bar{l_2} \textrm{ except } \\ &  k_1\!=\! k_2\!=\!\bar{k_1}\!=\!\bar{k_2}, l_1\!=\! l_2\!=\!\bar{l_1}\!=\!\bar{l_2} \\
				\hspace{-8pt}|E\{x^2[k_1,l_1]\}|^2 \qquad & k_1\!=\!\bar{k_2}, k_2\!=\!\bar{k_1}, l_1\!=\!\bar{l_2}, l_2\!=\!\bar{l_1} \textrm{ except } \\ &  k_1\!=\! k_2\!=\!\bar{k_1}\!=\!\bar{k_2}, l_1\!=\! l_2\!=\!\bar{l_1}\!=\!\bar{l_2} \\
				0 & \textrm{Otherwise}
			\end{array}
			\right.
		\end{align}
		Then, $E\{X[n_1,m_1]X^{*}[n_2,m_2]X^{*}[\bar{n_1},\bar{m_1}]X[\bar{n_2},\bar{m_2}]\}$ can be computed using \eqref{eq_2} and \eqref{eq_app9} as \eqref{eq_app10}, given at the top of the page. 
		
		\setcounter{equation}{17}
		\begin{figure*}			
			\begin{align}
				E\{X[n_1,m_1]X^{*}[n_2,m_2]X^{*}[\bar{n_1},\bar{m_1}]X[\bar{n_2},\bar{m_2}]\} \!=\!& E\{|x[k,l]|^4\} \biggl[NC_2\bigg(\dfrac{n_1\!-\! n_2\!-\!\bar{n_1}\!+\!\bar{n_2}}{N}\bigg) MC_2\bigg(\dfrac{m_1\!-\! m_2\!-\!\bar{m_1}\!+\!\bar{m_2}}{M}\bigg)\biggr] + \nonumber \\ &|E\{|x[k,l]|^2\}|^2 \biggl[- 2 NC_2\bigg(\dfrac{n_1\!-\! n_2\!-\!\bar{n_1}\!+\!\bar{n_2}}{N}\bigg) MC_2\bigg(\dfrac{m_1\!-\! m_2\!-\!\bar{m_1}\!+\!\bar{m_2}}{M}\bigg)+\nonumber \\ &\hspace{10pt} NC_2\bigg(\dfrac{n_1\!-\! n_2}{N}\bigg) NC_2\bigg(\dfrac{\bar{n_1}\!-\!\bar{n_2}}{N}\bigg) MC_2\bigg(\dfrac{m_1\!-\! m_2}{M}\bigg)MC_2\bigg(\dfrac{\bar{m_1}\!-\!\bar{m_2}}{M}\bigg) + \nonumber \\ &\hspace{10pt} NC_2\bigg(\dfrac{n_1\!-\!\bar{n_1}}{N}\bigg) NC_2\bigg(\dfrac{n_2\!-\!\bar{n_2}}{N}\bigg)MC_2\bigg(\dfrac{m_1\!-\!\bar{m_1}}{M}\bigg) MC_2\bigg(\dfrac{m_2\!-\!\bar{m_2}}{M}\bigg) \biggr] + \nonumber \\ &|E\{x^2[k,l]\}|^2 \biggl[- NC_2\bigg(\dfrac{n_1\!-\! n_2\!-\!\bar{n_1}\!+\!\bar{n_2}}{N}\bigg) MC_2\bigg(\dfrac{m_1\!-\! m_2\!-\!\bar{m_1}\!+\!\bar{m_2}}{M}\bigg) + \nonumber \\ &\hspace{10pt} NC_2\bigg(\dfrac{n_1\!+\!\bar{n_2}}{N}\bigg)  NC_2\bigg(\dfrac{n_2\!+\!\bar{n_1}}{N}\bigg) MC_2\bigg(\dfrac{m_1\!+\!\bar{m_2}}{M}\bigg)MC_2\bigg(\dfrac{m_2\!+\!\bar{m_1}}{M}\bigg) \biggr].	\label{eq_app11} 	
			\end{align}
			\hrule
			\begin{align}
				E\{A^2(\tau,f_d)\} &\!=\! |A_g\big(-\tau,f_d\big)|^2 \bigg(\sum_{n_1=0}^{N-1}\sum_{\bar{n_1}=0}^{N-1}\textrm{e}^{j2\pi [n_1-\bar{n_1}]Tf_d}\bigg) \bigg(\sum_{m_1=0}^{M-1}\sum_{\bar{m_1}=0}^{M-1}\textrm{e}^{j2\pi (m_1-\bar{m_1})\Delta f\tau}\bigg) \nonumber \\ &
				+\sum_{n_1=0}^{N-1}\sum_{n_2=0}^{N-1}\sum_{m_1=0}^{M-1}\sum_{m_2=0}^{M-1}	\bigg|A_g\bigg([n_1-n_2]T-\tau,[m_1-m_2]\Delta f+f_d\bigg)\bigg|^2 \nonumber \\ &
				+ \sum_{n_1=0}^{N-1}\sum_{n_2=0}^{N-1}\sum_{m_1=0}^{M-1}\sum_{m_2=0}^{M-1}	\sum_{\bar{n_1}=0}^{N-1}\sum_{\bar{n_2}=0}^{N-1} \sum_{\bar{m_1}=0}^{M-1}\sum_{\bar{m_2}=0}^{M-1}\textrm{e}^{j\pi (m_1+m_2-\bar{m_1}-\bar{m_2})\Delta f\tau}\textrm{e}^{j\pi [n_1+n_2-\bar{n_1}-\bar{n_2}]Tf_d} \nonumber \\ & \, 
				\times A_g\bigg([n_1-n_2]T-\tau,[m_1-m_2]\Delta f+f_d\bigg)  A_g^{*}\bigg([\bar{n_1}-\bar{n_2}]T-\tau,[\bar{m_1}-\bar{m_2}]\Delta f+f_d\bigg)\nonumber \\ & 
				\times \biggl\{\!\dfrac{1}{NM}C_2\bigg(\dfrac{n_1\!-\! n_2\!-\!\bar{n_1}\!+\!\bar{n_2}}{N}\bigg) C_2\bigg(\dfrac{m_1\!-\! m_2\!-\!\bar{m_1}\!+\!\bar{m_2}}{M}\bigg)\!\!\biggl[E\{|x[k,l]|^4\}\!-\!|E\{x^2[k,l]\}|^2\!-\!2\biggr] \nonumber \\ &\hspace{10pt} + |E\{x^2[k,l]\}|^2 C_2\bigg(\dfrac{n_1+\bar{n_2}}{N}\bigg)C_2\bigg(\dfrac{n_2+\bar{n_1}}{N}\bigg)C_2\bigg(\dfrac{m_1+\bar{m_2}}{M}\bigg)C_2\bigg(\dfrac{m_2+\bar{m_1}}{M}\bigg) \biggr\}. \label{eq_app13} 
			\end{align}
			\hrule
		\begin{align}
			E\{\hat{A}(\tau,f_d)\} &\!=\! \dfrac{1}{NM}\!\!\sum_{n_1=0}^{N-1}\sum_{n_2=0}^{N-1}\sum_{m_1=0}^{M-1}\sum_{m_2=0}^{M-1}\biggl[\! E\{X[n_1,m_1]X^{*}[n_2,m_2]\} \textrm{e}^{j\pi (m_1+m_2)\Delta f\tau}\textrm{e}^{j\pi [(n_1+n_2+1)T+\tau]f_d} \nonumber \\ & \hspace{150pt} \times A_g\bigg([n_1-n_2]T-\tau,[m_1-m_2]\Delta f+f_d\bigg)\biggr]. \label{eq_app1} 
		\end{align}
		\hrule
		\setcounter{equation}{21}
		\begin{align}
			E\{X[n_1,m_1]X^{*}[n_2,m_2]\} &=  \textrm{e}^{j\pi\frac{(N-1)(n_1-n_2)}{N}}\dfrac{\sin(\pi (n_1-n_2))}{\sin(\pi (n_1-n_2)/N)}\textrm{e}^{-j\pi\frac{(m_1-m_2)(M-1)}{M}}\dfrac{\sin(\pi (m_1-m_2))}{\sin(\pi (m_1-m_2)/M)}. \label{eq_app4} 
		\end{align}
		\hrule
		\end{figure*}
		Note that $\sum_{k=0}^{N-1} \textrm{e}^{j2\pi kn/N} = N$ when $n/N \in \mathbb{Z}$. It can also be shown that $\sum_{k=0}^{N-1} \textrm{e}^{j2\pi kn/N}$ is zero when $\forall n \in \mathbb{Z}, n/N \not\in \mathbb{Z},$ using \cite[eq. (26)]{12967}. Therefore, by considering the integer nature of $n_1,n_2,\bar{n_1},\bar{n_2},m_1,m_2,\bar{m_1}$ and $\bar{m_2}$, we can simplify \eqref{eq_app10} as \eqref{eq_app11}, given at the next page, where $C_2\big(x)$ is given in \eqref{eq_C2}. Given that $0\!\le n_1,n_2 \le\! N-1$, we can further show that $(n_1\!-\! n_2)/N \!\in\!\mathbb{Z}$ only when $n_1\!=\! n_2$. Therefore, $C_2\big([n_1\!-\! n_2]/N\big)=1$ if $n_1=n_2$ and zero otherwise. This holds true for any $C_2\big([i-j]/N\big)$ such that $0\le i,j\le N-1$ and $C_2\big([i-j]/M\big)$ such that $0\le i,j\le M-1$. Thus, substituting \eqref{eq_app11} in \eqref{eq_app7} along with normalized symbol energy gives \eqref{eq_app13}, given at the next page. Finally, we use \cite[eq. (26)]{12967} to show $\sum_{n_1=0}^{N-1}\sum_{\bar{n_1}=0}^{N-1}\textrm{e}^{j2\pi [n_1-\bar{n_1}]x} = N^2C_1(x,N)$ where $C_1(x,N)$ is given in \eqref{eq_C1}. Substituting this in \eqref{eq_app13} completes the derivation of \eqref{eq_16}.
			
		\section{}\label{appendixB}	
		Let us start by computing the expectation of the complex AF as \eqref{eq_app1}, given at the next page. Given the i.i.d structure of $x[k,l]$, we note that $E\{x[k_1,l_1]x^{*}[k_2,l_2]\}$ is non-zero only when $k_1=k_2$ and $l_1=l_2$. Therefore, using \eqref{eq_2} and the normalized symbol energy, we can compute $E\{X[n_1,m_1]X^{*}[n_2,m_2]\}$ as 
		\setcounter{equation}{20}
		\begin{align}
			&E\{X[n_1,m_1]X^{*}[n_2,m_2]\} \nonumber \\ &\hspace{45pt}= \sum_{k_1=0}^{N-1} \textrm{e}^{j2\pi\frac{(n_1-n_2)k_1}{N}}\sum_{l_1=0}^{M-1}\textrm{e}^{-j2\pi \frac{(m_1-m_2)l_1}{M}}. \label{eq_app3} 
		\end{align} 
		Using \cite[eq. (26)]{12967} and following some straightforward mathematical manipulations, we can further simplify \eqref{eq_app3} as \eqref{eq_app4}, given at the next page. Please note that $|(n_1\!-\! n_2)/N|<1$ given $0\!\le n_1,n_2 \le\! N \!-\!1$. Therefore, with the integer nature of $n_1$ and $n_2$, the division of $\sin(\pi (n_1\!-\! n_2))$ and $\sin(\pi (n_1\!-\! n_2)/N)$ is $N$ when $n_1\!=\! n_2$ and zero otherwise. Similarly, the division between $\sin(\pi (m_1-m_2))$ and $\sin(\pi (m_1-m_2)/M)$ is $M$ when $m_1=m_2$ and zero otherwise. This makes
		\setcounter{equation}{22}
		\begin{align}
		E\{X[n_1,m_1]X^{*}[n_2,m_2]\} = \left\{
			\begin{array}{ll}
				MN \qquad & n_1\!=\!n_2, m_1\!=\!m_2 \\
				0 & \textrm{otherwise}.
			\end{array}
			\right. \label{eq_app5} 
		\end{align}
		Substituting \eqref{eq_app5} in \eqref{eq_app1} and further simplifying using \cite[eq. (26)]{12967}, we write
		\begin{align}
			|E\{\hat{A}(\tau,f_d)\}| &\!=\! NM\biggr[\bigg|\dfrac{\textrm{sinc}(\pi NTf_d)\textrm{sinc}(\pi M\Delta f\tau)}{\textrm{sinc}(\pi Tf_d)\textrm{sinc}(\pi\Delta f\tau)}\bigg| \nonumber \\ & \hspace{90pt}\times |A_g\big(-\tau,f_d\big)|\biggr]. \label{eq_app6} 
		\end{align}
		Substituting $A_g\big(-\tau,f_d\big)$ using \eqref{eq_9} and further simplifying when $\tau=0$ and $Tf_d\in \mathbb{Z}$ completes the derivation of \eqref{eq_15}.
		
	\end{appendices}
	

\end{document}